\documentclass{llncs}
\usepackage[draft]{botex}
\usepackage{verbatim}
\usepackage{tikz}
\usetikzlibrary{shapes,decorations}
\usepackage{prettyref}
\usepackage{algorithmic}
\usepackage{algorithm}

\newrefformat{cond}{Condition \ref{#1}}

\newcommand{\tsys}{\ensuremath{\mathcal{T}}}
\newcommand{\seq}{\ensuremath{\sigma}}
\newcommand{\sched}{\ensuremath{S}}
\newcommand{\backlog}{\ensuremath{\mathbf{B}}}
\newcommand{\backplus}{\ensuremath{\mathbf{B^+}}}
\newcommand{\supp}{\ensuremath{\mathrm{supp}}}
\newcommand{\conp}{\ccconp}

\title{Feasibility analysis of sporadic real-time multiprocessor task systems} 
\author{Vincenzo Bonifaci\inst{1} \and Alberto Marchetti-Spaccamela\inst{2}}
\institute{%
Max-Planck Institut f\"ur Informatik, Saarbr\"ucken, Germany \\
\email{bonifaci@mpi-inf.mpg.de}
\and
Sapienza Universit\`a di Roma, Rome, Italy \\
\email{alberto@dis.uniroma1.it}}

\begin{document}
\maketitle
\begin{abstract}
We give the first algorithm for testing the feasibility of a system of sporadic real-time tasks on a set of identical processors, solving one major open problem in the area of multiprocessor real-time scheduling \cite{Baruah:2009:open}. We also investigate the related notion of schedulability and a notion that we call online feasibility. Finally, we show that discrete-time schedules are as powerful as continuous-time schedules, which answers another open question in the above mentioned survey. 
\end{abstract}

\section{Introduction}

As embedded microprocessors become more and more common, so does the need to design systems that are guaranteed to meet deadlines in applications that are safety critical, where missing a deadline might have severe consequences. In such a real-time system, several tasks may need to be executed on a multiprocessor platform and a scheduling policy needs to decide which tasks should be active in which intervals, so as to guarantee that all deadlines are met. 

The \emph{sporadic task model} is a model of recurrent processes in hard real-time systems that has received great attention in the last years (see for example \cite{Baker:2007,Baruah:2009:open} and references therein). A sporadic task $\tau_i = (C_i, D_i, P_i)$ is characterized by a worst-case compute time $C_i$, a relative deadline $D_i$, and a minimum interarrival separation $P_i$. Such a sporadic task generates a potentially infinite sequence of jobs: each job arrives at an unpredictable time, after the minimum separation $P_i$ from the last job of the same task has elapsed; it has an execution requirement less than or equal to $C_i$ and a deadline that occurs $D_i$ time units after its arrival time. 
A sporadic task system $\tsys$ is a collection of such sporadic tasks. Since the actual interarrival times can vary, there are infinitely many job sequences that can be generated by $\tsys$. 

We are interested in designing algorithms that tell us when a given sporadic task system can be feasibly scheduled, with preemption and migration, on a set of $m \ge 1$ identical processors. The question can be formulated in several ways: 
\begin{itemize}
\item \emph{Feasibility}: is it possible to feasibly schedule on $m$ processors any job sequence that can be generated by \tsys?  
\item \emph{Online feasibility}: is there an online algorithm that can feasibly schedule on $m$ processors any job sequence that can be generated by \tsys? 
\item \emph{Schedulability}: does the given online algorithm $\mathrm{Alg}$ feasibly schedule on $m$ processors any job sequence that can be generated by \tsys? 
\end{itemize}

\paragraph{Previous work.}
Most of the previous work in the context of sporadic real-time feasibility testing has focused on the case of a single processor \cite{Baruah:2003}. The seminal paper by Liu and Layland  \cite{Liu:1973} gave a best possible fixed priority algorithm for the case where deadlines equal periods (a fixed priority algorithm initially orders the tasks and then -- at each time instant -- schedules the available job with highest priority).  
It is also known that  the Earliest Deadline First (EDF) algorithm, that schedules at any time the job with the earliest absolute deadline, is optimal in the sense that for any sequence of jobs it produces a valid schedule whenever a valid schedule exists \cite{Dertouzos:1974}. Because EDF is an online algorithm, this implies that the three questions of feasibility, of online feasibility and of schedulability with respect to EDF are equivalent. It was known for some time that EDF-schedulability could be tested in exponential time and more precisely that the problem is in \conp\ \cite{Baruah:1990}. 
The above results triggered a significant research effort within the scheduling community and many results have been proposed for specific algorithms and/or special cases; nonetheless, we remark that the feasibility problem for a single processor remained open for a long time and that only recently it has been proved \conp-complete \cite{Eisenbrand:2010}. 

The case of multiple processors is far from being as well understood as the single processor case. For starters, EDF is no longer optimal -- it is not hard to construct feasible task systems for which EDF fails, as soon as $m \ge 2$.  Another important difference with the single processor case is that here clairvoyance does help the scheduling algorithm: there exists a task system that is feasible, but for which no online algorithm can produce a feasible schedule on every job sequence \cite{Fisher:2009}. Thus, the notions of feasibility and on-line feasibility are distinct.  

On the positive side there are many results for special cases of the problem; however we remark that \emph{no optimal scheduling algorithm is known, and no test -- of whatsoever complexity -- is known} that correctly decides the feasibility or the online feasibility of a task system. This holds also for \emph{constrained-deadline} systems, in which deadlines do not exceed periods. The question of designing such a test has been listed as one of the main algorithmic problems in real-time scheduling \cite{Baruah:2009:open}.


Regarding schedulability, many schedulability tests are known for specific algorithms (see \cite{Baker:2007} and references therein), but, to the best of our knowledge, the only general test available is a test that requires exponential space  \cite{Baker:2007:b}.  


\paragraph{Our results.}
We study the three above problems in the context of constrained-deadline multiprocessor systems and we provide new  results for each of them.

First, for the feasibility problem, we give the first correct test, thus solving \cite[Open Problem 3]{Baruah:2009:open} for constrained-deadline systems. The test has high complexity, but it has the interesting consequence that a job sequence that witnesses the infeasibility of a task system \tsys\ has without loss of generality length at most doubly exponential in the bitsize of \tsys. 

Then we give the first correct test for the online feasibility problem. The test has exponential time complexity and is constructive: if a system is deemed online feasible, then an optimal online algorithm can be constructed (in the same time bound). Moreover, this optimal algorithm is without loss of generality \emph{memoryless}: its decisions depend only on the current (finite) state and not on the entire history up to the decision point (see Section \ref{sec:definitions} for a formal definition). 
These results suggest that the two problems of feasibility and online feasibility might have different complexity.

For the schedulability problem, we provide a general schedulability test showing that the schedulability of a system by any memoryless algorithm can be tested in polynomial space. This improves the result of Baker and Cirinei \cite{Baker:2007:b} (that provided an exponential space test for essentially the same class of algorithms).

We finally consider the issue of discrete time schedules versus continuous time schedules. The above results are derived with the assumption that the time line is divided into indivisible time slots and preemptions can occur only at integral points, that is, the schedule has to be discrete. In a continuous schedule, time is not divided into discrete quanta and preemptions may occur at any time instant. We show that in a sporadic task system a discrete schedule exists whenever a continuous schedule does, thus showing that the discrete time assumption is without loss of generality. Such  equivalence is known for periodic task systems (i.e. task system in which each job of a task is released exactly after the period $P_i$ of the task has elapsed); however, the reduction does not extend to the sporadic case and the problem is cited among the important open problems in real-time scheduling  \cite[Open Problem 5]{Baruah:2009:open}. 

All our results can be extended to the arbitrary-deadline case, at the expense of increasing some of the complexity bounds. In this extended abstract we restrict to the constrained-deadline case to simplify the exposition. 

Our main conceptual contribution is to show how the feasibility problem, the online feasibility problem and the schedulability problem can be cast as the problem of deciding the winner in certain two-player games of infinite duration played on a finite graph. We then use tools from the theory of games to decide who has a winning strategy. In particular, in the case of the feasibility problem we have a game of imperfect information where one of the players does not see the moves of the opponent, a so-called \emph{blindfold game} \cite{Reif:1984}. This can be reformulated as a one-player (i.e., solitaire) game on an exponentially larger graph and then solved via a reachability algorithm. However, a technical complication is that in our model a job sequence and a schedule can both have infinite length, which when the system is feasible makes the construction of a feasible schedule challenging. We solve this complication by an application of K\"onig's Infinity Lemma from graph theory \cite{Diestel:2005}. This is the technical ingredient that, roughly speaking, allows us to reduce the infinite job  sequences with infinite  length  to finite sequences and ultimately to obtain the equivalence between continuous and discrete schedules. 

The power of our new approach is its generality: it can be applied to all three problems and -- surprisingly -- it yields proofs that are not technically too complicated. 
We hope that this approach might be useful to answer similar questions for other real-time scheduling problems. 


\paragraph{Organization.} The remainder of the paper is structured as follows. In Section \ref{sec:definitions} we formally define the model and set up some common notation. In Section \ref{sec:algorithms} we describe and analyze our algorithms for feasibility and schedulability analysis. The equivalence between continuous and discrete schedules is treated in Section \ref{sec:continuous}, and we finish with some concluding remarks in Section \ref{sec:conclusion}. 

\section{Definitions}
\label{sec:definitions}

Let $\Nat = \{0,1,2,\ldots\}$ and $[n]=\{1,2,\ldots,n\}$. 
Given a set $X$, with $\binom{X}{k}$ we denote the set of all $k$-subsets of $X$. 

Consider a task system \tsys\ with $n$ tasks, and $m$ processors; without loss of generality, $m \le n$. Each task $i$ is described by three parameters: a worst-case \emph{compute time} $C_i$, a \emph{relative deadline} $D_i$, and a \emph{minimum interarrival time} $P_i$. We assume these parameters to be positive integers and that $D_i \le P_i$ for all $i$.  

\newcommand{\rct}{\ensuremath{\mathbf{C}}}
\newcommand{\ttd}{\ensuremath{\mathbf{D}}}
\newcommand{\tta}{\ensuremath{\mathbf{P}}}
\newcommand{\zero}{\ensuremath{\mathbf{0}}}
\newcommand{\scheddec}{\ensuremath{\mathbf{S}}}
\newcommand{\alg}{\ensuremath{\mathrm{Alg}}}

Let $\rct := \times_{i=1}^n ([C_i] \cup \{0\})$, $\ttd := \times_{i=1}^n ([D_i] \cup \{0\})$, $\tta := \times_{i=1}^n ([P_i] \cup \{0\})$, $\zero := (0)_{i=1}^n$. 
A \emph{job sequence} is a function $\seq: \Nat\ \to \rct$. The interpretation is that $\seq(t) = (\seq_i(t))_{i=1}^n$ iff, for each $i$ with $\seq_i(t)>0$, a new job from task $i$ is released at time $t$ with compute time $\seq_i(t)$, and no new job from task $i$ is released if $\seq_i(t)=0$. A \emph{legal} job sequence has the additional property that for any distinct $t,t' \in \Nat$ and any $i$, if $\seq_i(t)>0$ and $\seq_i(t') > 0$, then $|t-t'| \ge P_i$. A job sequence is \emph{finite} if $\seq(t')=\zero$ for all $t'$ greater or equal to some $t \in \Nat$; in this case, we say that the sequence has \emph{length} $t$. 

Let $\scheddec:=\cup_{k=0}^m \binom{[n]}{k}$. A \emph{schedule} is a function $\sched: \Nat\ \to \scheddec$; we interpret $\sched(t)$ as the set of those $k$ tasks ($0 \le k \le m$) that are being processed from time $t$ to time $t+1$ \footnote{Since $D_i \le P_i$, there can be at most one pending job from task $i$. In the arbitrary-deadline case, this can be generalized by considering $O(D_i/P_i)$ jobs.}. We allow that $\sched(t)$ contains a task $i$ even when there is no pending job from $i$ at time $t$; in that case there is no effect (this is formalized below). 

A \emph{backlog configuration} is an element of $\backlog := \rct \times \ttd \times \tta$. 
At time $t$, a backlog configuration $(c_i,d_i,p_i)_{i=1}^n \in \backlog$ \footnote{For notational convenience, here we have reordered the variables so as to have $n$-tuples of triples, instead of triples of $n$-tuples.} will denote the following: 
\begin{itemize}
\item $c_i \in [C_i] \cup \{0\}$ is the \emph{remaining compute time} of the unique pending job from task $i$, if any; if there is no pending job from task $i$, then $c_i=0$; 
\item $d_i \in [D_i] \cup \{0\}$ is the \emph{remaining time to deadline} of the unique pending job from task $i$, if any; if there is no pending job from task $i$, or the deadline has already passed, then $d_i=0$; 
\item $p_i \in [P_i] \cup \{0\}$ is the \emph{minimum remaining time to the next activation} of task $i$, that is, the minimum $p_i$ such that a new job from task $i$ could be legally released at time $t+p_i$. 
\end{itemize}
A configuration $(c_i,d_i,p_i)_{i=1}^n \in \backlog$ is a \emph{failure configuration} if for some task $i$, $c_i > 0$ and $d_i = 0$.  
\begin{remark}
\label{rmk:state-size}
The set $\backlog$ is finite, and its size is $2^{O(s)}$, where $s$ is the input size of $\tsys$ (number of bits in its binary encoding). 
\end{remark}

Given a legal job sequence $\seq$ and a schedule \sched, we define in the natural way an infinite sequence of backlog configurations $\langle \seq,\sched \rangle := b_0 b_1 \ldots$. The initial configuration is $b_0 := (0,0,0)_{i=1}^n$, and given a backlog configuration $b_t = (c_i,d_i,p_i)_{i=1}^n$, its successor configuration $b_{t+1} = (c'_i,d'_i,p'_i)_{i=1}^n$ is obtained as follows: 
\begin{itemize}
\item if $\seq_i(t) > 0$, then $c'_i = \seq_i(t) - x_i$, where $x_i$ is 1 if $i \in \sched(t)$, and 0 otherwise; moreover, $d'_i = D_i$ and $p'_i = P_i$; 
\item if $\seq_i(t) = 0$, then $c'_i = \max(c_i - x_i,0)$, where $x_i$ is defined as above; moreover, $d'_i = \max(d_i - 1,0)$ and $p'_i = \max(p_i - 1,0)$.  
\end{itemize}
We can now define a schedule $S$ to be \emph{feasible for} $\seq$ if no failure configuration appears in $\langle \seq, \sched \rangle$. Finally, a task system $\tsys$ is \emph{feasible} when every legal job sequence admits a feasible schedule. Stated otherwise, a task system is not feasible when there is a legal job sequence for which no schedule is feasible. We call such a job sequence a \emph{witness} of infeasibility. 

A \emph{deterministic online algorithm} \alg\ is a sequence of functions: 
$$ \alg_t : \rct^{t+1} \to \scheddec, \qquad t = 0,1,2,\ldots$$
By applying an algorithm $\alg$ to a job sequence $\seq$, one obtains the schedule $S$ defined by $S(t) = \alg_t( \seq(0), \ldots, \seq(t) ). $ Then $\alg$ feasibly schedules $\seq$ whenever $S$ does. 
\newcommand{\malg}{\ensuremath{\mathrm{Malg}}}
A \emph{memoryless} algorithm is a single function $\malg: \backlog \times \rct \to \scheddec$; it is a special case of an online algorithm in which the scheduling decisions at time $t$ are based only on the current backlog configuration and on the tasks that have been activated at time $t$. 

Finally, a task system \tsys\ is \emph{online feasible} if there is a deterministic online algorithm \alg\ such that every legal job sequence from \tsys\ is feasibly scheduled by \alg. We then say that \alg\  is  \emph{optimal} for \tsys, and that \tsys\ is \emph{schedulable} by \alg.  
Online feasibility implies feasibility, but the converse fails: there is a task system that is feasible, but that does not admit any optimal online algorithm \cite{Fisher:2009}. 

\section{Algorithms for feasibility and schedulability analysis}
\label{sec:algorithms}

\subsection{Feasibility}
\label{sec:feas}
We first model the process of scheduling a task system as a game between two players over infinitely many rounds. At round $t=0,1,2,\ldots$, the first player (the ``adversary'') selects a certain set of tasks to be activated. Then the second player (acting as the scheduler) selects a set of tasks to be processed, and so on. The game is won by the first player if a failure configuration is eventually reached. 

In order to capture the definition of feasibility correctly, the game must proceed so that the adversary has no information at all on the moves of the scheduler; in other words, the job sequence must be constructed obliviously from the schedule. This is because \emph{if the task system is infeasible, then a single witness job sequence must fail all possible schedules simultaneously}. 
Models of such games, where the first player has no information on the moves of the opponent, have been studied in the literature under the name of \emph{blindfold games} \cite{Reif:1984}. One approach to solving these games is to construct a larger one-player game, in which each state encodes all positions that are compatible with at least one sequence of moves for the second player. 

Given a task system \tsys, we build a bipartite graph $G^+(\tsys)=(V_1,V_2,A)$. Nodes in $V_1$ ($V_2$) will correspond to decision points for the adversary (scheduler). A node in $V_1$ or $V_2$ will encode mainly two kinds of information: (1) the counters that determine time to deadlines and next earliest arrival dates; and (2) the set of all plausible remaining compute times of the scheduler. 

Let $\backplus := \ttd \times \tta \times 2^{\rct}.$ Each of $V_1$ and $V_2$ is a copy of $\backplus$, so each node of $V_1$ is identified by a distinct element from $\backplus$, and similarly for $V_2$. 
We now specify the arcs of $G^+(\tsys)$. Consider an arbitrary node $v_1 \in V_1$ and let $((d_i,p_i)_{i=1}^n,Q)$ be its identifier, where $Q \in 2^\rct$. Its successors in $G^+(\tsys)$ are all nodes $v_2=((d'_i,p'_i)_{i=1}^n,Q') \in V_2$ for which there is a tuple $(k_i)_{i=1}^n \in \rct$ such that: 
\begin{enumerate}
\item $p_i = 0$ for all $i \in \supp(k)$, where $\supp(k)=\{i : k_i > 0\}$ (this ensures that each task in $k$ can be activated); 
\item $p'_i = P_i$, and $d'_i = D_i$ for all $i \in \supp(k)$ (activated jobs cannot be reactivated before $P_i$ time units); 
\item $p'_i = p_i$ and $d'_i = d_i$ for all $i \notin \supp(k)$ (counters of other tasks are not affected); 
\item \label{cond:complete1} each $(c'_i)_{i=1}^n \in Q'$ is obtained from some $(c_i)_{i=1}^n \in Q$ in the following way: $c'_i = k_i$ for all $i \in \supp(k)$, and $c'_i = c_i$ for all $i \notin \supp(k)$ (in every possible scheduler state, the remaining compute time of each activated job is set to the one prescribed by $k$);    
\item $Q'$ contains all $(c'_i)_{i=1}^n$ that satisfy \prettyref{cond:complete1}.    
\end{enumerate}
Now consider an arbitrary node $v_2 \in V_2$, say $v_2 = ((d_i,p_i)_{i=1}^n, Q)$. The only successor of $v_2$ will be the unique node $v_1=((d'_i,p'_i)_{i=1}^n, Q') \in V_1$ such that: 
\begin{enumerate}
\item $d'_i = \max(d_i-1,0)$, $p'_i = \max(p_i-1,0)$ for all $i \in [n]$ (this models a ``clock-tick''); 
\item \label{cond:complete2} for each $(c'_i)_{i=1}^n \in Q'$, there are an element $(c_i)_{i=1}^n \in Q$ and some $S \in \scheddec$ such that $c'_i = \max(c_i-1,0)$ for all $i \in S$ and $c'_i = c_i$ for all $i \notin S$ (each new possible state of the scheduler is obtained from some old state after the processing of at most $m$ tasks); 
\item \label{cond:complete3}
 for each $(c'_i)_{i=1}^n \in Q'$, one has, for all $i$, $c'_i = 0$ whenever $d'_i=0$ (this ensures that the resulting scheduler state is valid); 
\item $Q'$ contains all $(c'_i)_{i=1}^n$ that satisfy \prettyref{cond:complete2} and \prettyref{cond:complete3}. 
\end{enumerate}
That is, the only successor to $v_2$ is obtained by applying all possible decisions by the scheduler and then taking $Q'$ to be the set of all possible (valid) resulting scheduler states. Notice that because we only keep the valid states (\prettyref{cond:complete3}), the set $Q'$ might be empty. In this case we say that the node $v_1$ is a \emph{failure state}; it corresponds to some deadline having been violated. Also notice that any legal job sequence $\seq$ induces an alternating walk in the bipartite graph $G^+(\tsys)$ whose $(2t+1)$-th arc corresponds to $\seq(t)$.    

Finally, the \emph{initial state} is the node $v_0 \in V_1$ for which $d_i=p_i=0$ for all $i$, and for which the only possible scheduler state is \zero. (See \prettyref{fig:state-graph} in the Appendix for a partial illustration of the construction in the case of the task system $\tsys=((1,2,2), (2,2,2))$, for $m=1$.) Note that, given two nodes of $G^+(\tsys)$, it is easy to check their adjacency, in time polynomial in $|\backlog^+|$. 

\newcommand{\valid}{\ensuremath{\mathrm{valid}}}
\begin{definition}
For a legal job sequence \seq, the set of \emph{possible valid scheduler states} at time $t$ is the set of all $(c_i)_{i=1}^n \in \rct$ for which there exists a schedule $\sched$ such that (i) $\langle \seq, \sched \rangle = b_0 b_1 b_2 \ldots$ with no configuration $b_0,b_1,\ldots,b_t$ being a failure configuration, and (ii) the first component of $b_t$ is $(c_i)_{i=1}^n$. We denote this set by $\valid(\seq, t)$. 
\end{definition}

\begin{lemma}
\label{lem:graph-step}
Let $t \ge 0$ and let $((d_i,p_i)_{i=1}^n,Q) \in V_1$ be the node reached by following for $2t$ steps the walk induced by \seq\ in the graph $G^+(\tsys)$. Then $Q=\valid(\seq, t)$. 
\end{lemma}
\begin{proof}[sketch]
By induction on $t$. When $t=0$ the claim is true because the only possible scheduler state is the \zero\ state. For larger $t$ it follows from how we defined the successor relation in $G^+(\tsys)$ (see in particular the definition of $Q'$). 
\qed
\end{proof}

\begin{lemma}
\label{lem:reachability}
Task system \tsys\ is infeasible if and only if, in the graph $G^+(\tsys)$, some failure state is reachable from the initial state.  
\end{lemma}
\begin{proof}[sketch]
If there is a path from the initial state to some failure state, by \prettyref{lem:graph-step} we obtain a legal job sequence \seq\ that witnesses that for some $t$, $\valid(\seq,t) = \emptyset$, that is, there is no valid scheduler state for \seq\ at time $t$; so there cannot be any feasible schedule for \seq. 

Conversely, if no failure state is reachable from the initial state, for any legal job sequence \seq\ one has $\valid(\seq,t) \neq \emptyset$ for all $t$ by \prettyref{lem:graph-step}. This immediately implies that no \emph{finite} job sequence can be a witness of infeasibility. We also need to exclude witnesses of infinite length. To do this, we apply K\"onig's Infinity Lemma \cite[Lemma 8.1.2]{Diestel:2005} (also stated in the Appendix). Consider the infinite walk induced by \seq\ in $G^+(\tsys)$ and the corresponding infinite sequence of nonempty sets of possible valid scheduler states $Q^{0}, Q^{1}, \ldots$, where $Q^t := \valid(\seq,t)$. Each scheduler state $q \in Q^{t}$ ($t \ge 1$) has been derived by some scheduler state in $q' \in Q^{t-1}$ and so $q$ and $q'$ can be thought of as neighbors in an infinite graph on the disjoint union of $Q^0, Q^1, \ldots$ (see Figure \ref{fig:koenig-valid} in the Appendix). Then K\"onig's Lemma implies that there is a sequence $q_0 q_1 \ldots$ (with $q_t \in Q^t$) such that for all $t \ge 1$, $q_t$ is a neighbor of $q_{t-1}$. This sequence defines a feasible schedule for $\seq$.
\qed 
\end{proof}


\begin{theorem}
\label{thm:main}
The feasibility problem for a sporadic constrained-deadline task system $\tsys$ can be solved in time $2^{2^{O(s)}}$, where $s$ is the input size of $\tsys$. Moreover, if \tsys\ is infeasible, there is a witness job sequence of length at most $2^{2^{O(s)}}$. 
\end{theorem}
\begin{proof}
The graph has $2|\backlog^+|=2^{2^{O(s)}}$ nodes, so the first part follows from Lemma \ref{lem:reachability} and the existence of linear-time algorithms for the reachability problem. The second part follows similarly from the fact that the witness sequence $\seq$ can be defined by taking $\seq(t)$ as the set of task activations corresponding to the $(2t+1)$-th arc on the path from the initial state to the reachable failure state.  
\qed
\end{proof}

\newcommand{\reach}{\textsc{Reach}}

\begin{algorithm}[t]
\caption{Algorithm for the feasibility problem}
\label{alg:feasibility}
\begin{algorithmic}
\FORALL{failure states $v_f \in V_1$}
\IF{\reach($v_0$, $v_f$, $2|\backlog^+|$)} \RETURN \textbf{infeasible} \ENDIF
\ENDFOR 
\RETURN \textbf{feasible}
\end{algorithmic}
\end{algorithm}
\begin{algorithm}[t]
\caption{\reach($x$, $y$, $k$)}
\label{alg:reach}
\begin{algorithmic}
\IF{$k=0$} \RETURN \textbf{true} if $x=y$, \textbf{false} if $x \neq y$ \ENDIF
\IF{$k=1$} \RETURN \textbf{true} if $(x,y) \in A$, \textbf{false} otherwise \ENDIF
\FORALL{$z \in V_1 \cup V_2$}
\IF{\reach($x$, $z$, $\floor{k/2}$) \textbf{and} \reach($z$, $y$, $\ceil{k/2}$)}
	\RETURN\textbf{true}
\ENDIF
\ENDFOR
\RETURN\textbf{false}
\end{algorithmic}
\end{algorithm}

We can in fact improve exponentially the amount of memory needed for the computation. The idea is to compute the state graph as needed, instead of storing it explicitly (Algorithm \ref{alg:feasibility}). We enumerate all failure nodes; for each failure node $v_f$, we check whether there exists a path from $v_0$ to $v_f$ in $G^+(\tsys)$ by calling the subroutine \reach\ (Algorithm \ref{alg:reach}). This subroutine checks recursively whether there is a path from $x$ to $y$ of length at most $k$ by trying all possible midpoints $z$. Some readers might recognize that \reach\ is nothing but Savitch's reachability algorithm \cite{Savitch:1970}. This yields the following improvement. 

\begin{theorem}
The feasibility problem for a sporadic constrained-deadline task system $\tsys$ can be solved in space $2^{O(s)}$, where $s$ is the input size of $\tsys$. 
\end{theorem}
\begin{proof}
Any activation of Algorithm \ref{alg:reach} needs $O(\log |\backlog^+|)=2^{O(s)}$ space, and the depth of the recursion is at most $O(\log |\backlog^+|)=2^{O(s)}$.
\qed
\end{proof}

\subsection{Online feasibility}
\label{sec:onlinefeas}
An issue with the notion of feasibility as studied in the previous section is  that, when the task system turns out to be feasible, one is still left clueless as to how the system should be scheduled. The definition of online feasibility (see Section \ref{sec:definitions}) addresses this issue. 
It could be argued from a system design point of view that one should focus on the notion of online feasibility, rather than on the notion of feasibility. 
In this section we discuss an algorithm for testing online feasibility. 
  


The idea is again to interpret the process as a game between the environment and the scheduler, with the difference that now the adversary can observe the current state of the scheduler (the remaining compute times). In other words, the game is no longer a blindfold game but a perfect-information game. We construct a graph $G(\tsys)=(V_1,V_2,A)$ where $V_1=\backlog$ and $V_2=\backlog \times \rct$. The nodes in $V_1$ are decision points for the adversary (with different outgoing arcs corresponding to different tasks being activated) and the nodes in $V_2$ are decision points for the scheduler (different outgoing arcs corresponding to different sets of tasks being scheduled). There is an arc $(v_1,v_2) \in A$ if $v_2=(v_1,k)$ for some tuple $k=(k_i)_{i=1}^n \in \rct$ of jobs that can legally be released when the backlog configuration is $v_1$; notice the crucial fact that whether some tuple $k$ can legally be released can be decided on the basis of the backlog configuration $v_1$ alone. There is an arc $(v_2,v'_1)$ if $v_2=(v_1,k)$ and $v'_1$ is a backlog configuration that can be obtained by $v_1$ after scheduling some subset of tasks; again this depends only on $v_1$ and $k$. In the interest of space we omit the formal description of the adjacency relation (it can be found in the Appendix). 

The game is now played with the adversary starting first in state $b_0=(0,0,0)_{i=1}^n$. The two players take turns alternately and move from state to state by picking an outgoing arc from each state. The adversary wins if it can reach a state in $V_1$ corresponding to a failure configuration. The scheduler wins if it can prolong play indefinitely while never incurring in a failure configuration. 
\begin{lemma}
\label{lem:onlinefeas}
The first player has a winning strategy in the above game on $G(\tsys)$ if and only if \tsys\ is not online feasible. Moreover, if \tsys\ is online feasible, then it admits an optimal memoryless deterministic online algorithm. 
\end{lemma}
\begin{proof}[sketch]
If the first player has a winning strategy $s$, then for any online algorithm \alg, the walk in $G(\tsys)$ obtained when player 1 plays according to $s$ and player 2 plays according to \alg, ends up in a failure configuration. But then the job sequence corresponding to this walk in the graph (given by the odd-numbered arcs in the walk) defines a legal job sequence that is not feasibly scheduled by \alg. 

If, on the other hand, the first player does not have a winning strategy, from the theory of two-player perfect-information games it is known (see for example \cite{Graedel:2002,McNaughton:1993}) that the second player has a winning strategy and that this can be assumed to be, without loss of generality, a deterministic strategy that depends only on the current state in $V_2$ (a so-called memoryless, or positional, strategy). Hence, for each node in $V_2$ it is possible to remove all but one outgoing arc so that in the remaining graph no failure configuration is reachable from $b_0$. The set of remaining arcs that leave $V_2$ implicitly defines a function from $V_2=\backlog \times \rct$ to $\scheddec$, that is, a memoryless online algorithm, which feasibly schedules every legal job sequence of \tsys.
\qed  
\end{proof}

\begin{theorem}
The online feasibility problem for a sporadic constrained-deadline task system $\tsys$ can be solved in time $2^{O(s)}$, where $s$ is the input size of $\tsys$. If \tsys\ is online feasible, an optimal memoryless deterministic online algorithm for \tsys\ can be constructed within the same time bound. 
\end{theorem}
\begin{proof}
We first construct $G(\tsys)$ in time polynomial in $|\backlog \times (\backlog \times \rct)| = 2^{O(s)}$. We then apply the following inductive algorithm to compute the set of nodes $W \subseteq V_1 \cup V_2$ from which player 1 can force a win; its correctness has been proved before (see for example \cite[Proposition 2.18]{Graedel:2002}). Define the set $W_i$ as the set of nodes from which player 1 can force a win in at most $i$ moves, so $W = \cup_{i \ge 0} W_i$. The set $W_0$ is simply the set of all failure configurations. The set $W_{i+1}$ is computed from $W_i$ as follows: 
\begin{align*}
W_{i+1} = W_i &\cup \{ v_1 \in V_1: (v_1,w) \in A \text{ for some } w \in W_i\} \\
& \cup \{ v_2 \in V_2: w \in W_i \text{ for all } (v_2,w) \in A \}. 
\end{align*}
At any iteration either $W_{i+1} = W_i$ (and then $W = W_i$) or $W_{i+1} \setminus W_i$ contains at least one node. Since there are $2^{O(s)}$ nodes, this means that $W = W_k$ for some $k = 2^{O(s)}$. Because every iteration can be carried out in time $2^{O(s)}$, it follows that the set $W$ can be computed within time $(2^{O(s)})^2 = 2^{O(s)}$. By \prettyref{lem:onlinefeas}, \tsys\ is online feasible if and only if $b_0 \notin W$. 

The second part of the claim follows from the second part of \prettyref{lem:onlinefeas} and from the fact that a memoryless winning strategy for player 2 (that is, an optimal memoryless scheduler) can be obtained by selecting, for each node $v_2 \in V_2 \setminus W$, any outgoing arc that does not have an endpoint in $W$.  
\qed
\end{proof}

\subsection{Schedulability}


In the case of the schedulability problem, we observe that the construction of Section \ref{sec:feas} can be applied in a simplified form, because for every node of the graph there is now \emph{at most one} possible valid scheduler state, which can be determined by querying the scheduling algorithm. 
This implies that the size of the graph reduces to $2|\backlog|=2^{O(s)}$. By applying the same approach as in Section \ref{sec:feas}, we obtain the following. 

\begin{theorem}
The schedulability problem for a sporadic constrained-deadline task system $\tsys$ can be solved in time $2^{O(s^2)}$ and space $O(s^2)$, where $s$ is the input size of $\tsys$. 
\end{theorem}
\begin{proof}[sketch]
Any activation of Algorithm \ref{alg:reach} needs $O(\log |\backlog|)=O(s)$ space, and the depth of the recursion is at most $O(\log |\backlog|)=O(s)$, so in total a space of $O(s^2)$ is enough. The running time can be found by the recurrence $T(k) = 2^{O(s)} \cdot 2 \cdot T(k/2) + O(1) $ which gives $T(k) = 2^{O(s \log k)}$ and finally $T(2|\backlog|)=2^{O(s^2)}$. 
\qed
\end{proof}


\section{Continuous versus discrete schedules}
\label{sec:continuous}
In this section we show that, under our assumption of integer arrival times for the jobs, the feasibility of a sporadic task system does not depend on whether one is considering discrete or continuous schedules. 

Let $J$ be the (possibly infinite) set of jobs generated by a job sequence $\seq$. In this section we do not need to keep track of which tasks generate the jobs, so it will be convenient to use a somewhat different notation. Let $r_j$, $c_j$, $d_j$ denote respectively the release date, compute time and absolute deadline of a job $j$; so job $j$ has to receive $c_j$ units of processing in the interval $[r_j,d_j]$. A \emph{continuous schedule} for $J$ on $m$ processors is a function $w : J \times \Nat \to \Real_+$ such that:
\begin{enumerate}
\item $w(j,t) \le 1$ for all $j \in J$ and $t \in \Nat$;
\item $\sum_{j \in J} w(j,t) \le m$ for all $t \in \Nat$.   
\end{enumerate}
Quantity $w(j,t)$ is to be interpreted as the total amount of processing dedicated to job $j$ during interval $[t,t+1]$. Thus, the first condition forbids the parallel execution of a job on more than one processor; the second condition limits the total volume processed in the interval by the $m$ processors. 
The continuous schedule $w$ is \emph{feasible} for \seq\ if it additionally satisfies 
\begin{enumerate}
\item[3.] $\sum_{r_j \le t < d_j} w(j,t) \ge c_j$ for all $j \in J$.
\end{enumerate}

Finally, a task system \tsys\ is \emph{feasible with respect to continuous schedules} if any legal job sequence \seq\ from \tsys\ has a feasible continuous schedule. For the sake of clarity we call a system that is feasible in the sense defined in Section \ref{sec:definitions} \emph{feasible with respect to discrete schedules}. 

\begin{theorem}
\label{thm:continuous}
A sporadic constrained-deadline task system \tsys\ is feasible with respect to continuous schedules iff it is feasible with respect to discrete schedules. 
\end{theorem}
\begin{proof}
If a task system is feasible with respect to discrete schedules, it is obviously also feasible with respect to continuous schedules: a discrete schedule is just a special case of a continuous schedule where $w(j,t) \in \{0,1\}$. So assume that a task system \tsys\ is feasible with respect to continuous schedules, but not with respect to discrete schedules. Then there must be a witness job sequence \seq\ that cannot be scheduled by any discrete schedule, but can be scheduled by some continuous schedule. 
By Theorem \ref{thm:main}, we can assume that \seq\ has some finite length $L \le 2^{2^{O(|\tsys|)}}$. So it generates a \emph{finite} collection of jobs $J$. But any feasible continuous schedule for a finite collection of jobs can be converted into a feasible discrete schedule \cite{Baruah:1996,Baruah:1990,Horn:1974} (see Appendix for a self-contained proof). This contradicts the initial assumption. 
\qed
\end{proof}

\section{Conclusion}
\label{sec:conclusion}
We have given upper bounds on the complexity of testing the feasibility, the online feasibility, and the schedulability of a sporadic task system on a set of identical processors. It is known that these three problems are at least \conp-hard \cite{Eisenbrand:2010}; however, no sharper hardness result is known. A natural question is to characterize more precisely the complexity of these problems, either by improving on the algorithms given here, or by showing that these problems are hard for some complexity class above \conp. 


\emph{Acknowledgments.}
We thank Sanjoy K.\ Baruah, Nicole Megow and Sebastian Stiller for useful discussions. 
\vspace{-0.3cm}

\bibliography{journals-algo,bonifaci,complexity,game-theory,real-time}

\providecommand\SortNoop[1]{}
\begin{thebibliography}{10}

\bibitem{Baker:2007}
T.~P. Baker and S.~K. Baruah.
\newblock Schedulability analysis of multiprocessor sporadic task systems.
\newblock In S.~H. Son, I.~Lee, and J.~Y.-T. Leung, editors, {\em Handbook of
  Real-Time and Embedded Systems}, chapter~3. CRC Press, 2007.

\bibitem{Baker:2007:b}
T.~P. Baker and M.~Cirinei.
\newblock Brute-force determination of multiprocessor schedulability for sets
  of sporadic hard-deadline tasks.
\newblock In {\em Proc. of 11th Conf. on Principles of Distributed Systems},
  pages 62--75, 2007.

\bibitem{Baruah:1996}
S.~K. Baruah, N.~K. Cohen, C.~G. Plaxton, and D.~A. Varvel.
\newblock Proportionate progress: A notion of fairness in resource allocation.
\newblock {\em Algorithmica}, 15(6):600--625, 1996.

\bibitem{Baruah:2003}
S.~K. Baruah and J.~Goossens.
\newblock Scheduling real-time tasks: Algorithms and complexity.
\newblock In J.~Y.-T. Leung, editor, {\em Handbook of Scheduling: Algorithms,
  Models, and Performance Analysis}, chapter~28. CRC Press, 2003.

\bibitem{Baruah:2009:open}
S.~K. Baruah and K.~Pruhs.
\newblock Open problems in real-time scheduling.
\newblock {\em Journal of Scheduling}, 2009.
\newblock doi:10.1007/s10951-009-0137-5.

\bibitem{Baruah:1990}
S.~K. Baruah, L.~E. Rosier, and R.~R. Howell.
\newblock Algorithms and complexity concerning the preemptive scheduling of
  periodic, real-time tasks on one processor.
\newblock {\em Real-Time Systems}, 2(4):301--324, 1990.

\bibitem{Dertouzos:1974}
M.~L. Dertouzos.
\newblock Control robotics: The procedural control of physical processes.
\newblock In {\em Proc. IFIP Congress}, pages 807--813, 1974.

\bibitem{Diestel:2005}
R.~Diestel.
\newblock {\em Graph theory}.
\newblock Springer, Heidelberg, 3rd edition, 2005.

\bibitem{Eisenbrand:2010}
F.~Eisenbrand and T.~Rothvo{\ss}.
\newblock {EDF}-schedulability of synchronous periodic task systems is
  {coNP}-hard.
\newblock In {\em Proc. 21st Symp. on Discrete Algorithms}, pages 1029--1034,
  2010.

\bibitem{Fisher:2009}
N.~Fisher, J.~Goossens, and S.~K. Baruah.
\newblock Optimal online multiprocessor scheduling of sporadic real-time tasks
  is impossible.
\newblock Technical Report 09--009, University of North Carolina at Chapel
  Hill, Department of Computer Science, Chapel Hill, NC, 2009.

\bibitem{Graedel:2002}
E.~Gr{\"a}del, W.~Thomas, and T.~Wilke, editors.
\newblock {\em Automata, Logics, and Infinite Games: A Guide to Current
  Research}, volume 2500 of {\em Lecture Notes in Computer Science}. Springer,
  2002.

\bibitem{Horn:1974}
W.~A. Horn.
\newblock Some simple scheduling algorithms.
\newblock {\em Naval Research Logistics Quarterly}, 21:177--185, 1974.

\bibitem{Liu:1973}
C.~L. Liu and J.~W. Layland.
\newblock Scheduling algorithms for multiprogramming in a hard-real-time
  environment.
\newblock {\em Journal of the ACM}, 20(1):46--61, 1973.

\bibitem{McNaughton:1993}
R.~McNaughton.
\newblock Infinite games played on finite graphs.
\newblock {\em Annals of Pure and Applied Logic}, 65(2):149--184, 1993.

\bibitem{Reif:1984}
J.~H. Reif.
\newblock The complexity of two-player games of incomplete information.
\newblock {\em Journal of Computer and System Sciences}, 29(2):274--301, 1984.

\bibitem{Savitch:1970}
W.~J. Savitch.
\newblock Relationships between nondeterministic and deterministic tape
  complexities.
\newblock {\em Journal of Computer and Systems Sciences}, 4(2):177--192, 1970.

\end{thebibliography}
\bibliographystyle{abbrv}

\newpage
\appendix
\section{Appendix}

\subsection{K\"onig's Infinity Lemma}

A \emph{ray} is an infinite graph $(V, E)$ of the form
$$ V = \{x_0,x_1,x_2,...\}	,\,\, E = \{(x_0, x_1), (x_1, x_2), (x_2, x_3), \ldots\}. $$

\spnewtheorem*{koenig}{K\"onig's Infinity Lemma}{\rmfamily\bfseries}{\itshape}

\begin{koenig}
Let	$Q^0, Q^1, \ldots$ be an infinite sequence of disjoint nonempty finite sets of nodes, and let $G$ be a graph on their union. Assume that every node $q$ in a set $Q^t$ with $n \ge 1$ has a predecessor $q'$ in $Q^{t-1}$, so that $(q', q)$ is an arc of $G$. Then $G$ contains a ray $q_0, q_1, \ldots $ with $q_t \in Q^t$ for all $t$. 
\end{koenig}
\begin{proof}
See for example \cite[Lemma 8.1.2]{Diestel:2005} (the result is stated there in terms of undirected graphs, but the proof works equally well for the directed case.)  
\qed
\end{proof}

\subsection{Definition of $G(\tsys)$ in Section \ref{sec:onlinefeas}}
Recall that $\backlog = \rct \times \ttd \times \tta$ and that $G(\tsys)=(V_1,V_2,A)$, where $V_1 = \backlog$ and $V_2 = \backlog \times \rct$. 
We specify the adjacency relation $A$. Consider an arbitrary node $v_1 \in V_1$ with $v_1=(c_i,d_i,p_i)_{i=1}^n$. Its successors in $G(\tsys)$ are all nodes $v_2=((c'_i,d'_i,p'_i)_{i=1}^n,(k_i)_{i=1}^n) \in V_2$ with $(k_i)_{i=1}^n \in \rct$ such that: 
\begin{enumerate}
\item $p_i = 0$ for all $i \in \supp(k)$, where $\supp(k)=\{i : k_i > 0\}$ (this ensures that each task in $k$ can be activated); 
\item $p'_i = P_i$, and $d'_i = D_i$ for all $i \in \supp(k)$ (activated jobs cannot be reactivated before $P_i$ time units); 
\item $p'_i = p_i$ and $d'_i = d_i$ for all $i \notin \supp(k)$ (counters of other tasks are not affected); 
\item $c'_i = k_i$ for all $i \in \supp(k)$, and $c'_i = c_i$ for all $i \notin \supp(k)$ (the remaining compute time of each activated job is set to the one prescribed by $k$);    
\end{enumerate}
Now consider an arbitrary node $v_2 \in V_2$, say $v_2 = ((c_i,d_i,p_i)_{i=1}^n, (k_i)_{i=1}^n)$. Its successors in $G(\tsys)$ are all nodes $v_1=((c'_i,d'_i,p'_i)_{i=1}^n) \in V_1$ for which there is some $S \in \scheddec$ such that: 
\begin{enumerate}
\item $d'_i = \max(d_i-1,0)$, $p'_i = \max(p_i-1,0)$ for all $i \in [n]$ (this models a ``clock-tick''); 
\item $c'_i = \max(c_i-1,0)$ for all $i \in S$ and $c'_i = c_i$ for all $i \notin S$ (the new remaining compute times are obtained from the old remaining compute times by processing at most $m$ tasks). 
\end{enumerate}


\subsection{Missing details for \prettyref{thm:continuous}}
\begin{proof}[Missing details for \prettyref{thm:continuous}]
We show that any finite job set $J$ has a discrete schedule whenever it has a continuous schedule. 
We setup an instance of a maximum flow problem whose solutions correspond to continuous schedules for \seq, and whose integral solutions correspond to discrete schedules for \seq; see also a similar construction in \cite{Baruah:1996,Baruah:1990,Horn:1974}.   
We build a network $N$ consisting of four types of nodes: 
\begin{enumerate}
\item a source node $a$; 
\item for every $t=0,1,\ldots,\max_{j \in J} d_j$, a node $x_t$; 
\item for every job $j$ in \seq, a node $q_j$; 
\item a sink node $z$.  
\end{enumerate}
The arcs of the network are as follows:
\begin{enumerate}
\item from $a$ to each Type 2 node, an arc with capacity $m$ ($m$ is the number of processors); 
\item from each Type 2 node $x_t$ to each Type 3 node $q_j$ such that $r_j \le t < d_j$, an arc with capacity 1; 
\item from each Type 3 node $q_j$ to $z$, an arc with capacity $c_j$.  
\end{enumerate}
Let $K$ be the sum of the capacities of Type 3 edges. 

Assume that a feasible continuous schedule $w$ exists for \seq. We now define a flow by setting the flow on each arc $(a,x_t)$ to be $\sum_{j \in J} w(j,t)$; the flow on each arc $(x_t,q_j)$ to $w(j,t)$; and the flow on each arc $(q_t,z)$ to $c_j$. Now conditions (1) and (2) in the definition of continuous schedule for $w$ ensure that the capacity constraints are satisfied. Condition (3) ensures that the amount of flow entering any Type 3 node $q_j$ is at least 
$$ \sum_{r_j \le t < d_j} w(j,t) \ge c_j. $$
Notice that some Type 3 node might have more flow entering the node than leaving it; but in that case we can still obtain a feasible flow of the same value by decreasing the incoming flow, and the capacities will not be violated.   

Since all the capacities are integral we know there must be an optimal integral flow; its value will be $K$. From this we can extract a discrete schedule by setting $w(j,t)$ equal to the flow on arc $(x_t,q_j)$; this value must be either 0 or 1 by integrality. The total flow collected at each node $q_j$ is exactly $c_j$. We obtain a feasible discrete schedule for \seq, contradicting our initial assumption.  
\qed
\end{proof}

\newcommand{\stext}[1]{\begin{minipage}{2.2cm}\centering #1\end{minipage}}
\begin{figure}
\begin{center}
\tikzstyle{my}=[circle split,draw=black,text centered]
\begin{tikzpicture}[->,thick,node distance=4 cm] 
\node (v11) [my,label={above left:$v_0$}] {\stext{$d_1$: 0, $t_1$: 0,\\ $d_2$: 0, $t_2$: 0} \nodepart{lower} \stext{$c_1$: 0, $c_2$: 0\\ \ \\I }}; 
\node (v21) [right of=v11,my] {\stext{$d_1$: 0, $t_1$: 0,\\ $d_2$: 0, $t_2$: 0} \nodepart{lower} \stext{$c_1$: 0, $c_2$: 0\\ \ \\II }}; 
\node (v24) [above of=v21,node distance=4cm,my] {\stext{$d_1$: 2, $t_1$: 2,\\ $d_2$: 0, $t_2$: 0} \nodepart{lower} \stext{$c_1$: 1, $c_2$: 0\\ \ \\II}}; 
\path (v11) edge (v24); 
\node (v14) [above of=v11] {...}; 
\path (v24) edge [bend left] (v14); 
\path (v11) edge [bend left] (v21);
\path (v21) edge [bend left] (v11);  
\node (v12) [below of=v11,my] {\stext{$d_1$: 1, $t_1$: 1,\\ $d_2$: 1, $t_2$: 1} \nodepart{lower} \stext{$c_1$: 0, $c_2$: 2\\ $c_1$: 1, $c_2$: 1\\ $c_1$: 1, $c_2$: 2\\I }}; 
\node (v22) [below of=v21,my] {\stext{$d_1$: 2, $t_1$: 2,\\ $d_2$: 2, $t_2$: 2} \nodepart{lower} \stext{$c_1$: 1, $c_2$: 2\\ \ \\II }}; 
\path (v11) edge (v22); 
\path (v22) edge [bend left] (v12); 
\node (v23) [below of=v22,my] {\stext{$d_1$: 1, $t_1$: 1,\\ $d_2$: 1, $t_2$: 1} \nodepart{lower} \stext{$c_1$: 0, $c_2$: 2\\ $c_1$: 1, $c_2$: 1\\ $c_1$: 1, $c_2$: 2\\II }}; 
\path (v12) edge (v23); 
\node (v13) [left of=v23,my] {\stext{$d_1$: 0, $t_1$: 0,\\ $d_2$: 0, $t_2$: 0} \nodepart{lower} \stext{$\emptyset$\\ \ \\I } }; 
\path (v23) edge [bend left] (v13); 
\node (vxx) [left of=v11,my] {\stext{$d_1$: 0, $t_1$: 0,\\ $d_2$: 2, $t_2$: 2} \nodepart{lower} \stext{$c_1$: 0, $c_2$: 2\\ \ \\II }}; 
\path (v11) edge [bend left] (vxx); 
\node (vyy) [above of=vxx] {...}; 
\path (vxx) edge [bend left] (vyy); 
%
\end{tikzpicture}
\caption{A subgraph of the graph $G^+(\tsys)$ for the task system $\tsys=((1,2,2),(2,2,2))$ and $m=1$. Nodes labeled with ``I'' are in $V_1$, nodes labeled with ``II'' are in $V_2$.}
\label{fig:state-graph}
\end{center}
\end{figure}
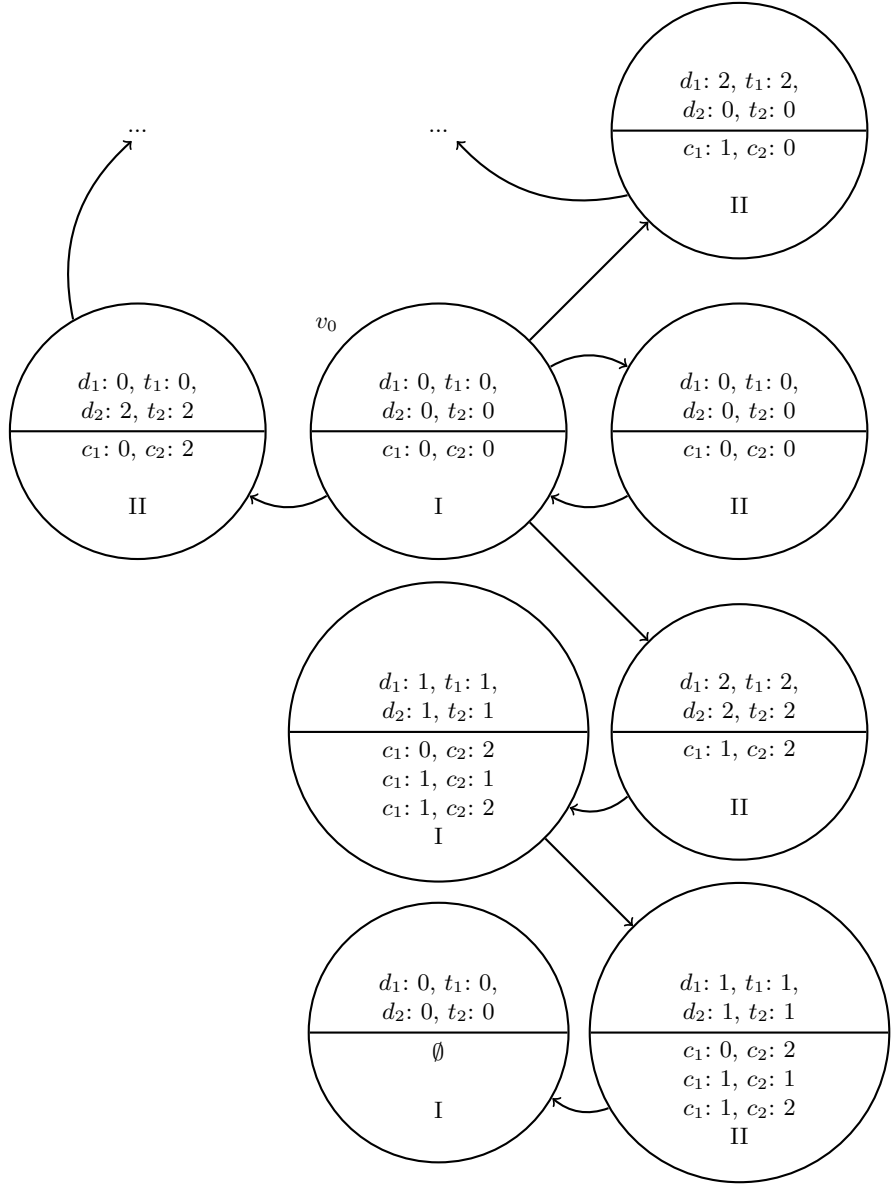

\begin{figure}
\begin{center}
\tikzstyle{my}=[thin,minimum size=2.25cm,circle split,draw=black,text centered]
\tikzstyle{vss}=[thin,circle,draw=black]
\begin{tikzpicture}[->,thick,node distance=3.25cm] 
\node (v1) [label={below left:$Q^0$},label={above left:$v_0$},my] {}; 
\node (v2) [label={below left:$Q^1$},my,right of=v1] {}; 
\node (v3) [label={below left:$Q^2$},my,right of=v2] {}; 
\node (v4) [label={below left:$Q^3$},my,right of=v3] {}; 
\node (v5) [right of=v4,node distance=1.8cm] {}; 
\path (v1) edge (v2); 
\path (v2) edge (v3); 
\path (v3) edge (v4); 
\node (vss1) [vss,below of=v1,node distance=0.5cm] {}; 
\node (vss21) [vss,below of=v2,node distance=0.25cm] {}; 
\node (vss22) [vss,below of=v2,node distance=0.75cm] {}; 
\node (vss31) [vss,below of=v3,node distance=0.20cm] {}; 
\node (vss32) [vss,below of=v3,node distance=0.55cm] {}; 
\node (vss33) [vss,below of=v3,node distance=0.90cm,pin={[pin edge={-}]-120:valid scheduler state}] {}; 
\node (vss41) [vss,below of=v4,node distance=0.25cm] {}; 
\node (vss42) [vss,below of=v4,node distance=0.75cm] {}; 
\path (vss1) edge [dashed] (vss21); 
\path (vss1) edge [] (vss22); 
\path (vss21) edge [dashed] (vss31); 
\path (vss22) edge [] (vss32); 
\path (vss22) edge [dashed] (vss33); 
\path (vss32) edge [] (vss41); 
\path (vss32) edge [dashed] (vss42); 
\node (vss51) [below of=v5,node distance=0.5cm] {}; 
\path (vss41) edge [dashed] (vss51); 
\draw (0,1.5) -- (9.5,1.5) node [pos=0.5,label={above:Walk in $G^+(\tsys)$ (nodes in $V_1$)}] { }; 
\end{tikzpicture}
\caption{Illustration of how K\"onig's Lemma applies to the proof of Lemma \ref{lem:reachability}. A prefix of an infinite ray is shown in solid lines.}
\label{fig:koenig-valid}
\end{center}
\end{figure}
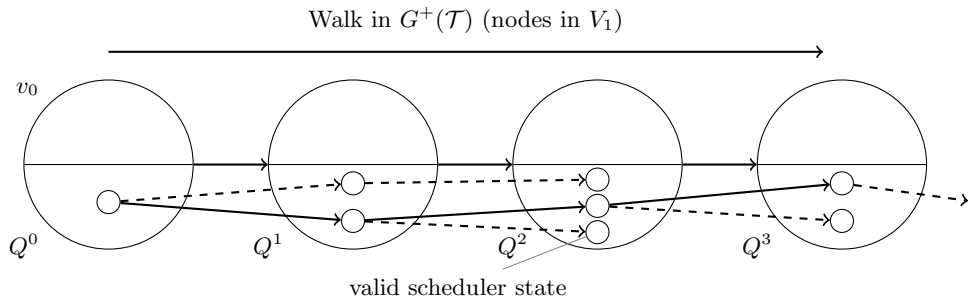

\end{document}